\documentstyle[pra,aps,twocolumn]{revtex}

\def\QED{\leavevmode\unskip\penalty9999 \hbox{}\nobreak\hfill
     \quad\hbox{\leavevmode  \hbox to.77778em{%
               \hfil\vrule   \vbox to.675em%
               {\hrule width.6em\vfil\hrule}\vrule\hfil}}
     \par\vskip3pt}

\def\ibb #1{\leavevmode\hbox{\kern.3em\vrule
     height 1.5ex depth -.1ex width .4pt\kern-.3em\rm#1}}
\def\Cx {{\ibb C}}
\def\Rx {{\ibb R}}

\title{Can spectral and local information decide separability?}
 \author{K.~G.~H. Vollbrecht\thanks{Electronic Mail: \tt{k.vollbrecht@tu-bs.de}}
 {{}\ and\ }M.~M. Wolf\thanks{Electronic Mail: \tt{mm.wolf@tu-bs.de}}
   \\[1ex]
  {\small Institut f{\"u}r Mathematische Physik, TU Braunschweig,}\\
  {\small Mendelssohnstr.3, 38106 Braunschweig, Germany.}}
\date{\today}

\begin{document}
\draft \maketitle

\begin{abstract}We discuss the discriminating power of separability
criteria, which are based on the spectrum of a quantum state and
its reductions. Common examples are entropic inequalities
utilizing conditional Tsallis or R\'enyi entropies. We prove that
these inequalities are implied by the reduction criterion for any
positive value of the entropic parameters. We show however, that
arbitrary sets of criteria based on spectral and local information
can never be sufficient by establishing a separable, isospectral
and locally undistinguishable counterpart for any Werner state in
odd dimensions. For the case of two qubit systems we show that a
simple controlled phase gate operation can produce an isospectral,
entangled state out of a separable one, which has the same
reductions.
\end{abstract}


\narrowtext

\section{Introduction}
Entanglement has always been a key issue in the ongoing debate
about the foundations and interpretation of quantum mechanics
since Einstein \cite{EPR} and Schr\"odinger \cite{Sch} expressed
their deep dissatisfaction about this astonishing part of quantum
theory. Whereas for the long period from 1935 to 1964, until Bell
\cite{Bell} published his famous work, discussions about
entanglement were purely meta-theoretical, nowadays quantum
information theory has established entanglement as a physical
resource and key ingredient for quantum computation and quantum
information processing. This new point of view led to a dramatic
increase of general structural knowledge about entanglement in the
last few years. However, there are still many open problems and
deciding them often fails at what we try to profit from in quantum
computation: the rapid increase of dimensions. One of these
unsolved problems is in fact the question whether a given quantum
state is entangled or merely classically correlated, i.e.
separable.

The present paper is devoted to the question to what extent the
combination of spectral and local information can decide
separability. We will show that the discriminating power of any
separability criterion based on this kind of information is rather
limited or even null for some otherwise simple classes of quantum
states.

The paper is organized as follows. In Sec. \ref{entropicsec} we
will discuss entropic inequalities as common examples for the kind
of criteria we want to address. We give a small counterexample for
the conjectured monotonicity of the conditional Tsallis entropy
and prove that the criteria based on conditional Tsallis and
R\'enyi entropies are implied by the reduction criterion for any
positive value of the entropic parameters. In Sec.
\ref{Wernersec}, \ref{qubitsec} we will construct pairs of {\it
isospectral} states having the same spectra and the same
reductions such that one state is entangled and the other is
separable. Sec. \ref{Wernersec} establishes such an isospectral
and separable counterpart for any entangled Werner state in odd
dimensions. Sec. \ref{qubitsec} then provides a similar example
even for the case of two qubits by utilizing a phase gate
operation showing clearly how weak the discriminating power of the
considered class of separability criteria is.

 To fix ideas we will start by
recalling some of the basic notions and previous results.

A bipartite quantum state described by its density matrix $\rho$
acting on a Hilbert space ${\cal H}={\cal H}^{(1)}\otimes{\cal
H}^{(2)}$ is said to be {\it separable}, unentangled or
classically correlated if it can be written as a convex
combination of tensor product states \cite{Werner89}
\begin{equation}\label{separable}
\rho = \sum_j p_j \rho_j^{(1)}\otimes\rho_j^{(2)},
\end{equation}
where the positive weights $p_j$ sum up to one and $\rho^{(i)}$
describes a state on ${\cal H}^{(i)}$. This means in particular
that pure states are separable if and only if they are product
states. Moreover, all entanglement properties of pure states,
which can always be written in their Schmidt form (cf.
\cite{Schmidt}) as $ |\Psi\rangle = \sum_i \sqrt{\lambda_i}
|i\rangle\otimes|i\rangle, $  are completely determined by the
eigenvalues $\{\lambda_i\}$ of the reduced state $\rho_A={\rm
tr}_B|\Psi\rangle\langle\Psi|$.

For mixed quantum states however, the situation is much more
difficult and deciding whether a state is entangled or separable
is not yet feasible in general. Currently, the most efficient
necessary criterion for separability is the positivity of the
partial transpose (PPT), i.e., the condition that $\rho^{T_1}$ has
to be a positive semi-definite operator \cite{Peres}. The partial
transpose of the state is thereby defined in terms of its matrix
elements with respect to some basis by $\langle
kl|\rho^{T_1}|mn\rangle=\langle ml|\rho|kn\rangle$. For the
smallest non-trivial systems with $2\times2$ resp. $2\times3$
dimensional Hilbert spaces and a few other special cases
 the PPT-criterion also turned out to be
sufficient \cite{HHH}. In higher dimensional systems however, so
called {\it bound entangled} states exist, which satisfy the
PPT-condition without being separable \cite{be}.

Another well known condition is given by the {\it reduction
criterion} \cite{red,secondred}
\begin{equation}\label{reduction}\rho_A\otimes{\bf
1}-\rho\geq 0,
\end{equation}which is implied by the PPT-criterion but
nevertheless an important condition since its violation implies
the possibility of recovering entanglement by distillation (which
is yet unclear for PPT violating states). For the case of two
qubits the reduction criterion is also known to be
 sufficient \cite{red}. The general line of implication is
$$
    \rho \mbox{ sep.} \;\Rightarrow\; \rho^{T_1}\geq 0\;
\Rightarrow\rho \mbox{ undistillable}\Rightarrow\rho_A\otimes{\bf
1}-\rho\geq 0.$$

\section{Entropic inequalities}\label{entropicsec}
The idea to use entropic inequalities as separability criteria for
mixed states goes back to the mid nineties when Cerf and Adami
\cite{Cerf} and the Horodecki family \cite{HHHalpha} recognized
that certain conditional R\'enyi entropies are non-negative for
separable states, and it was recently resurrected by several
groups \cite{Abe1,Abe2,Lloyd,Vidiella,Rajagopal} in the form of
conditional Tsallis entropies. Characteristic for these criteria
is, that they just use the spectra of the state and its
reductions.

The quantum R\'enyi entropy depending on the parameter
$\alpha\in\Rx$ is given by
\begin{equation}\label{Renyi}
S_\alpha (\rho) = \frac{\log {\rm tr}(\rho^\alpha)}{1-\alpha}.
\end{equation}
For the case of separable states it was shown in
\cite{Cerf,HHHalpha,rank}:
\begin{equation}\label{Renyiineq}
S_\alpha(\rho)-S_\alpha(\rho_A)\geq 0
\end{equation}for
$\alpha=0,\infty$ and $\alpha\in[1,2]$, where $S_0, S_1, S_\infty$
reduces to the logarithm of the rank, the von Neumann entropy and
the negative logarithm of the operator norm respectively.

In Ref. \cite{Abe1,Lloyd} essentially the same criterion was
expressed in terms of the conditional Tsallis entropy
\begin{equation}\label{Tsalliscond}
T_\alpha(\rho)=\frac{{\rm tr}(\rho_A^\alpha)-{\rm
tr}(\rho^\alpha)}{(\alpha-1)\;{\rm tr}(\rho_A^\alpha)}\; ,
\end{equation}
which was conjectured \cite{Lloyd} to be monotone in $\alpha$ and
non-negative for all positive values of $\alpha$ if and only if
$\rho$ is separable.

However, it is definitively neither monotone nor sufficient for
separability. Monotonicity can most easily be ruled out by low
rank examples like $$
\rho=\frac12\big(|\Phi_+\rangle\langle\Phi_+|+|01\rangle\langle
01|\big),\quad
|\Phi_+\rangle=\frac1{\sqrt{2}}\big(|00\rangle+|11\rangle\big), $$
for which the reduced state has eigenvalues $\frac14,\frac34$ and
therefore $T_0=T_\infty=0\neq T_2=\frac15$. Fortunately however,
monotonicity is not necessary for proving the validity of the
Tsallis/R\'enyi separability criteria for other values than
$\alpha=0,\infty$, $\alpha\in[1,2]$ \cite{wrongproof}:

\vspace{3pt}{\bf Proposition 1 }{\it Any state for which the
reduction criterion (\ref{reduction}) holds, in particular any
separable state, satisfies
\begin{equation}\label{essence}
  {\rm tr}(\rho_A^\alpha) -  {\rm tr}(\rho^\alpha)\;\left\{\begin{array}{cc}
    \geq 0\;, & \;\alpha > 1 \\
    \leq 0\;, & \;0\leq\alpha < 1 \
  \end{array}\right. .
\end{equation}This implies the validity of the
Tsallis/R\'enyi separability criteria for any positive value of
the entropic parameter.}\vspace{3pt}

{\it Proof : } For $\alpha > 1$ the proof is essentially based on
the Golden-Thompson inequality (cf.\cite{Petz}) stating that ${\rm
tr}\big(e^A e^B\big) \geq {\rm tr}\big( e^{A+B}\big)$ for
 hermitian matrices $A,B$. Utilizing the
definition of the reduced state, i.e., $\forall P\geq 0 : {\rm
tr}\big(\rho(P\otimes{\bf 1} )\big)\equiv {\rm tr}\big(\rho_A P
\big)$ this leads to:
\begin{eqnarray}
{\rm tr}\big(\rho_A^\alpha\big) &=& {\rm
tr}\big[\rho(\rho_A^{\alpha-1}\otimes{\bf 1})\big]\\ &=& {\rm
tr}\Big[\exp\big(\ln\rho\big)\exp\big((\alpha-1)\ln(\rho_A\otimes{\bf
1} )\big)\Big]\\ &\geq& {\rm
tr}\Big[\exp\big(\ln(\rho)+(\alpha-1)\ln(\rho_a\otimes{\bf 1}
)\big)\Big]\label{mono1}\\ &\geq& {\rm
tr}\Big[\exp\big(\alpha\ln\rho\big)\Big]= {\rm
tr}\big(\rho^\alpha\big)\label{mono2} ,
\end{eqnarray}
where Eq.(\ref{mono1})-(\ref{mono2}) is implied by the reduction
criterion (\ref{reduction}) since the logarithm is operator
monotone \cite{KL} and the exponential function is monotone under
the trace. The latter can be seen by noting that for any $A$
hermitian, $P\geq 0$ and $B=(A+\epsilon P)$ with $\epsilon\geq 0$:
\begin{equation}\label{exp}
\frac{\partial}{\partial\epsilon} {\rm tr}\big(e^B\big)={\rm
tr}\big(e^B P\big)\geq 0.
\end{equation}
Hence ${\rm tr}\big(e^B\big)\geq{\rm tr}\big(e^A\big)$ is implied
by $B\geq A$.

For $0\leq\alpha < 1$ the reduction criterion can immediately be
applied since $f(A)=A^r$ is an operator decreasing function for
$-1\leq r \leq 0,\; A\geq 0$ (cf.\cite{MO}) and thus
\begin{equation}\label{al1}
{\rm tr}\big(\rho_A^\alpha\big)={\rm
tr}\big[\rho(\rho_A^{\alpha-1}\otimes{\bf 1})\big]\leq {\rm
tr}\big(\rho^\alpha\big),
\end{equation}
which completes the proof.\QED

For negative values of $\alpha$, $T_\alpha(\rho)\geq 0$ holds for
any state no matter if it is entangled or not. This can be shown
by noting that for $\{|a\rangle \}$ being an eigenbasis of
$\rho_A$
 \begin{eqnarray}
 {\rm tr}\big(\rho^\alpha_A\big) &=& \sum_a \langle
 a|\rho_A|a\rangle^\alpha
 = \sum_a\Big[\sum_i\langle a\otimes i|\rho|a\otimes i\rangle\Big]^\alpha\label{f1}\\
 &\leq& \sum_{a,i}\langle a\otimes i|\rho|a\otimes i\rangle^\alpha\leq
{\rm tr}\big(\rho^\alpha\big) ,\label{f2}
 \end{eqnarray}
 where Eq.(\ref{f1}-\ref{f2}) uses that $\big(\sum_i b_i\big)^\alpha\leq\sum_i b_i^\alpha$ holds for $b_i\geq 0,\; \alpha\leq0$, and the last inequality is implied by the convexity of negative
 powers.

We note that the von Neumann case $S_1(\rho_A)\leq S_1(\rho)$
resp. $T_1(\rho)\geq 0$ also follows from the reduction criterion
 due to the operator monotonicity of the
logarithm. Hence, separability criteria based on conditional
Tsallis-R\'enyi entropies for all positive values of the entropic
parameter are implied by the reduction criterion and therefore
certainly weaker than the latter. In particular they cannot be
sufficient.

This raises the question about the resolution that separability
criteria, which are based on the spectrum of a state and its
reductions, can have in general. The next two sections are devoted
to the somewhat disillusioning answer.

\section{Separable counterparts for Werner
states}\label{Wernersec} Werner states \cite{Werner89} have always
played an important and paradigmatic role in quantum information
theory. Their characteristic property is that they commute with
all unitaries of the form $U\otimes U$ and they can be expressed
as
\begin{equation}\label{Werner}
\rho(p)= (1-p)\frac{P_+}{r_+} + p \frac{P_-}{r_-} , \quad 0\leq p
\leq 1,
\end{equation}
where $P_+$ ($P_-$) is the projector onto the symmetric
(antisymmetric) subspace of $\Cx^d\otimes\Cx^d$ and $r_\pm = {\rm
tr}[P_\pm] = \frac{d^2\pm d}2$ are the respective dimensions.
Werner showed that these states are entangled iff $p>\frac12$
independent of the dimension $d$. The following proposition shows
however, that none of these entangled states for odd dimension $d$
can be detected by any separability criterion, which is based on
the spectrum of the state and its reductions.

\vspace{3pt}{\bf Proposition 2 }{\it Any entangled state in
$\Cx^d\otimes\Cx^d$ with maximal chaotic reductions and
eigenvalues having multiplicities which are multiples of $d$, has
a separable isospectral counterpart, which is locally
undistinguishable as it has the same reductions. }\vspace{3pt}

{\it Proof: }  Let us consider a special basis of maximally
entangled states in $\Cx^d\otimes\Cx^d$ \cite{W2}:
\begin{equation}\label{mebasis}
|\Psi_{jk}\rangle=\frac{1}{\sqrt{d}}\sum_{n=1}^d
\exp\Big(\frac{2\pi i}{d} j n \Big)|n, n\oplus k\rangle,
\end{equation}where $j,k=1,\ldots,d$ and $\oplus$ means addition
modulo $d$. Any equal weight combination of all states of the form
(\ref{mebasis}), which belong to the same value of $k$, is then a
projector onto a separable state since
\begin{eqnarray}\label{Psep}
P_k&=&\sum_{j=1}^d |\Psi_{jk}\rangle\langle\Psi_{jk}|\\
&=&\frac{1}{d}\hspace{-3pt}\sum_{j,n,m=1}^d\hspace{-4pt}
\exp\Big[\frac{2\pi i}{d} j(n-m) \Big] |n, n\oplus k\rangle\langle
m, m\oplus k|\nonumber\\ &=& \sum_{n=1}^d |n\rangle\langle
n|\otimes |n\oplus k\rangle\langle n\oplus k|\nonumber
\end{eqnarray}is an equal weight combination of product states.
Here we have used that $\frac1d \sum_{j=1}^d \exp\Big[\frac{2\pi
i}{d} j(n-m)\Big]=\delta_{n,m}$. Moreover, the reductions of the
respective states $P_k/d$ are maximally chaotic, i.e. $\rho_A={\bf
1}/d$, just as the reductions of any maximally entangled state.

If we now have a state with multiplicities being multiples of $d$
we can replace the projectors onto its eigenspaces with
sufficiently many projectors of the form $P_k$. The resulting
state will then be again a convex combination of product states,
i.e., separable, having the same spectrum and maximal chaotic
reductions.\QED

For the case of Werner states we note that the unitary invariance
of the state $\rho(p)$ in Eq. (\ref{Werner}) implies that its
reductions are $\rho_A={\bf 1}/d$. Moreover $\rho(p)$ has two
eigenvalues $(1-p)/r_+$ and $ p/r_-$ with multiplicities $r_+$,
$r_-$ which are indeed multiples of $d$ in odd dimensions.

Following Proposition 2 we can now construct a state
\begin{equation}\label{isowerner}
\rho'(p)=\frac{(1-p)}{r_+}\sum_{k=1}^{r_+/d} P_k + \frac{ p
}{r_-}\sum_{l=1}^{r_-/d} P_{l+r_+/d} ,
\end{equation} which has then both, the same spectrum and the same
reductions as $\rho(p)$. However, as convex combination of
separable states it is itself separable for any $0\leq p\leq 1$.

\section{The two qubit case}\label{qubitsec}

For the case of two qubits the antisymmetric subspace is one
dimensional and the Werner state $\rho(p)$ is therefore entangled
iff it has an eigenvalue larger than one half. It was essentially
shown by Bennett et al. \cite{Bennett} that this is indeed
sufficient to decide separability for two qubit states with
maximal chaotic reductions, such that these states cannot have
separable counterparts.

However, the following will show that separability is in general
not encoded in the spectrum of a two qubit state and its
reductions.

It is often advantageous to express two qubit states in terms of
the matrix
\begin{equation}\label{Rmatrix}
R_{ij}={\rm tr}\big(\rho \sigma_i\otimes\sigma_j\big) , \quad
i,j=0,\ldots,3
\end{equation} with $\sigma_0={\bf 1}$ and $\sigma_{1,2,3}$ being the Pauli matrices.
Normalization then requires $R_{0,0}=1$ and the reductions are
determined by the vectors $R_{0,i}$ and $R_{i,0}$. However,
positivity of the respective state has to be verified separately.

Let us now consider a controlled phase gate operation described by
the unitary
\begin{eqnarray}\label{phasegate}
U &=& |0\rangle\langle 0|\otimes {\bf 1} + |1\rangle\langle
1|\otimes \sigma_3\\ &=& \frac12({\bf 1}+\sigma_3)\otimes {\bf
1}+\frac12({\bf 1}-\sigma_3)\otimes\sigma_3
\end{eqnarray}which applies a phase operation on the target bit if the
source bit is in the state $1$ and leaves it unchanged if the
source bit is in the state $0$. Controlled phase gates as well as
controlled NOT gates can be considered as universal basic building
blocks for future quantum computers since together with single
qubit gates they can be used to implement an arbitrary quantum
circuit (cf. \cite{universality}).

The operation $U$ interchanges the matrix elements
$R_{1,0}\leftrightarrow R_{1,3}$, $R_{0,1}\leftrightarrow R_{3,1}$
and $R_{1,1}\leftrightarrow R_{2,2}$
since\begin{eqnarray}\label{Uaction} U (\sigma_1\otimes {\bf 1})
U^* &=& (\sigma_1\otimes\sigma_3)\; ,\\U ({\bf 1}\otimes\sigma_1)
U^* &=& (\sigma_3\otimes\sigma_1)\; , \\ U (\sigma_1\otimes
\sigma_1) U^* &=& (\sigma_2\otimes\sigma_2)\; .
\end{eqnarray}If we now choose the state such that
$R_{0,1}=R_{1,0}=R_{3,1}=R_{1,3}=r$ take on the same values,
$R_{1,1}=\frac12$ and the others vanish, then the controlled phase
gate will leave the reductions unchanged and simply interchange
$R_{1,1}\leftrightarrow R_{2,2}$:
\begin{equation}\label{RR}
R=\left(\begin{tabular}{c|c c c }
  1 & r & 0 & 0 \\ \hline
  r & $\frac12$ & 0 & r \\
  0 & 0 & 0 & 0 \\
  0 & r & 0 & 0 \\
\end{tabular}\right)
  \;\stackrel{U}{\mapsto}\;
  R'=\left(\begin{tabular}{c|c c c }
  1 & r & 0 & 0 \\ \hline
  r & 0 & 0 & r \\
  0 & 0 & $\frac12$ & 0 \\
  0 & r & 0 & 0 \\
\end{tabular}\right)
\end{equation}
These matrices correspond to positive states for $|r|\leq
\frac38$. Moreover, since the partial transposition only changes
the signs of the $\sigma_2$ components, the initial state is equal
to its partial transpose and therefore separable on the whole
interval. This argumentation fails however for the state $\rho
'=U\rho U^*$, which corresponds to $R'$, and in fact the
determinant of its partial transpose is negative and the state is
thus entangled for $\frac{\sqrt{3}}8<|r|\leq\frac38$.

Hence, the controlled phase gate operation can produce an
entangled state out of a separable one, which has the same
reductions and due to the unitarity of the operation the same
spectrum.

It is worth mentioning that although $\rho'$ is entangled  for
$\frac{\sqrt{3}}8<|r|\leq\frac38$ it does not violate the CHSH
Bell inequality \cite{CHSH}. It can be shown that $T_2\geq 0$ is
indeed stronger than the Bell inequality for two qubit systems
\cite{HHHalpha}. Hence, a similar example with one state being
separable and the other violating the CHSH inequality cannot
exist. This is however no longer true for higher dimensional
systems.

\section{Conclusion}

Deciding whether a given quantum state is entangled or not is
still one of the big open problems in the theory of quantum
information. We discussed to what extent criteria based on the
spectrum of a state and its reductions can decide separability. We
showed that criteria, which are based on entropic inequalities
using conditional Tsallis and R\'enyi entropies, are definitely
weaker than the reduction criterion but nevertheless valid for any
 value of the entropic parameter. Moreover, the fact that
there exists a separable counterpart for any entangled Werner
state in odd dimensions shows that there is only a rather loose
relation between the spectra of a state and its reductions on the
one hand and its entanglement properties on the other.

Since the latter must not depend on the choice of local bases, it
is sufficient to look at local unitary invariants of a state in
order to decide its separability. Surprisingly, the most efficient
separability criterion is still based on just one rather simple
invariant: the smallest eigenvalue of the partial transpose. One
remaining question in this context would therefore be: how can
other (easy calculable) invariants provide information about the
separability of a state, which is not yet encoded in the smallest
eigenvalue of its partial transpose?

\section*{Acknowledgement}
The authors would like to thank R.F. Werner for useful
discussions. Funding by the European Union project EQUIP (contract
IST-1999-11053) and financial support from the DFG (Bonn) is
gratefully acknowledged.


\begin{references}
\bibitem{secondred}Note that throughout this paper we omit equations corresponding to
the reduction with respect to the second subsystem $\rho_B={\rm
tr}_A(\rho)$. However, all the equations hold in a symmetric form
for both reductions and the presented ones should be read as
representatives for either of the two.
\bibitem{wrongproof}Abe and Rajagopal already tried to prove this proposition for separable states in
\cite{Abe1}. However, they failed since they wrongly supposed that
all the states in the decomposition in Eq.(\ref{separable})
mutually commute.
\end{references}
\end{document}